\documentclass[preprint,showpacs,prl]{revtex4}

\usepackage{amsmath}
\usepackage{graphicx}
\usepackage{dcolumn}
\usepackage{bm}

\begin{document}

\title{Multiple Cotunneling in Large Quantum Dot Arrays}
\author{T.~B.~Tran$^{1}$, I.~S.~Beloborodov$^{2}$, X.~M.~Lin$^{2,3}$, V.~M.~Vinokur$^{2}$, and
H.~M.~Jaeger$^{1}$}
\address{$^{1}$James Franck Institute, University of Chicago, Chicago, Illinois 60637, USA \\
$^{2}$Materials Science and $^{3}$Chemistry Divisions, Argonne National Laboratory,
Argonne, Illinois 60439, USA}

\date{\today}

\begin{abstract}
We investigate the effects of inelastic cotunneling on the electronic transport properties of gold nanoparticle multilayers and thick films at low applied bias, inside the Coulomb blockade regime. We find that the zero-bias conductance, $g_0(T)$, in all systems exhibits  Efros-Shklovskii-type variable range hopping transport. The resulting typical hopping distance, corresponding to the number of tunnel junctions participating in cotunneling events, is shown to be directly related to the power law exponent in the measured current-voltage characteristics.  We discuss the implications of these findings in light of models on cotunneling and hopping transport in mesoscopic, granular conductors.
\end{abstract}

\maketitle

A signature of electronic transport in the dielectric state of disordered materials such as doped semiconductors or granular metals is an exponential decrease of conductance with decreasing temperature~\cite{Fritzsche, Efros}. In semiconductors the underlying transport mechanism is hopping conductivity: thermally assisted tunneling between dopant sites or traps. The interplay between tunneling and thermal activation processes results in variable range hopping (VRH) characterized by a stretched exponential temperature dependence of the conductance, with details depending on the strength of the electron-electron interactions. This picture is based on hopping between atomic impurity sites in an otherwise uniform host material.

There is a large, increasingly important class of materials, including quench-condensed metal films and arrays of metallic~\cite{Black, other_arrays, Parthasarathy} or semiconducting~\cite{Yu} quantum dots, which  consist of weakly-coupled nanometer-scale islands.  Experimental data~\cite{Black,Yu} from mesoscopically-disordered granular materials typically exhibit a temperature dependence of the zero-bias conductance, $g_0(T)$, that is well fit by the Efros-Shklovskii (ES) model for VRH in the presence of strong electron-electron interactions~\cite{Efros}. However, the question to what extent VRH concepts can be applied to these granular systems has long remained unresolved \cite{Fritzsche}. One problem is that VRH elemental processes in granular systems involve tunneling through mesoscopic islands. In addition, it is not obvious how the Coulomb correlation between the initial and final grains, which is the basis for the ES mechanism, can be established in the presence of screening by intermediate conducting islands. Alternative models based on nearest neighbor hopping can reproduce the experimentally observed temperature dependence for $g_0(T)$, but only for specific assumptions about the distributions of tunnel distances and island sizes~\cite{Sheng}.

Very recently a new theoretical approach based on the concept of {\it multiple cotunneling}~\cite{Beloborodov,Feigelman} was developed which can explain and quantitatively describe the ES-VRH-like behavior of $g_0(T)$ observed in large granular systems. Cotunneling allows for charge transport through several junctions at a time by cooperative electron motion. Single cotunneling, introduced by Ref~\cite{Averin} and previously observed in small systems comprised of 2-3 tunnel junctions ~\cite{Geerligs, cotunnel_expt}, provides a conduction channel at low applied biases, where otherwise Coulomb blockade would suppress current flow. Here we investigate the opposite, large array limit to directly test the connection between multiple cotunneling and the temperature dependence of the zero-bias conductance. 

Extended, highly-ordered layers of self-assembled semiconducting or metallic quantum dots offer unique opportunities to explore the mesoscopic analog to VRH transport. In this Letter we focus on multilayers formed from ligated gold nanocrystals. These systems are nearly defect-free and are not plagued by the inhomogeneities inherent to quench-condensed thin films. Prepared between in-plane electrodes, the layers' structural integrity can be characterized down to the level of individual dots by transmission electron microscopy. Compared to monolayers, multilayers have one key experimental advantage: their significantly greater lower-sheet conductance enhances cotunneling and allows for tracking of $g_0(T)$ over a much wider temperature range deep inside the Coulomb blockade regime.

In cotunneling, an electron tunnels from an initial to final state via virtual intermediate states. In principle, both elastic and inelastic cotunneling processes are possible. In elastic cotunneling, the same electron tunnels from one island to another through a virtual state on an intermediate grain, while for inelastic cotunneling separate electrons are involved that tunnel into and out of different states, creating electron-holes pairs \cite{Averin}. In quantum dot arrays, inelastic cotunneling processes occur via phonons or collisions with other electrons. Although both cotunneling processes predict ES-type conductivity at low temperatures ~\cite{Beloborodov}, only inelastic cotunneling gives rise to nonlinear current-voltage ($I-V$) curves ~\cite{Averin,Geerligs}. By tracking the temperature dependence of the zero-bias conductance together with the evolution of $I-V$ curves we show that inelastic cotunneling provides a fully consistent picture for the transport behavior of weakly coupled quantum dot arrays.

Dodecanthiol-ligated gold nanoparticles were synthesized by the digestive ripening method described in Ref.~\cite{Lin}. This provided particle diameters around 5.5nm with tight size control and dispersion less than 5 $\%$ in each batch. All films were prepared on 3mm $\times$ 4mm silicon substrates coated with 100nm amorphous silicon nitride ($Si_3N_4$). For sample characterization by transmission electron microscopy (TEM), these substrates contained 70$\mu$m $\times$ 70$\mu$m or 300$\mu$m $\times$ 300 $\mu$m ``windows'' areas under which the Si had been etched away to leave free standing, TEM-transparent $Si_3N_4$ membranes. Using electron beam lithography, planar chromium electrodes with gaps of 500-600nm, widths of 2-2.5 $\mu$m and thickness of 20nm were fabricated onto the substrates prior to film deposition. Multilayer films were assembled through a layer-by-layer deposition technique in which compact, highly-ordered monolayers were created at the water-air interface of a small water droplet~\cite{Eah}. The monolayers were picked up, similar to a Langmuir-Schaefer technique\cite{Roberts}, and added a layer at a time to the assembly already on the substrate. For comparison with published results on drop-deposited nanoparticle films tens to hundreds of particle diameters in thickness~\cite{Doty,Yu}, we also investigated thick films. To prepare these samples the substrate was heated to $85-95^0C$ on a hot plate, and 50-100 $\mu$l solution was deposited. During solvent evaporation large, three-dimensional crystals of nanoparticles formed over the windows. From characterization by scanning probe and scanning electron microscopy we obtained film thickness values in the range 3-4$\mu$m.

In total 15 samples (3 bilayers, 3 trilayers, 4 tetralayers, 5 thick films) were studied under similar conditions. $I-V$ measurements for all samples were performed over the range 10-160K, using a Keithley 6430 source meter and applying voltage biases from -20V to +20V at ramp rates of 0.04-0.10V/s. The magnitude of the applied bias was limited to 20V to avoid dielectric breakdown of the substrate material. TEM images were taken posterior to the transport measurements to preclude irradiation damage or contamination, even if undetectable in the images. While thick films ordinarily would be opaque to TEM, the highly ordered interior structure can be revealed by carefully dapping the sample with methanol to remove part of the film prior to TEM imaging.

Zoomed-in TEM images from the gap region between the electrodes are shown in Fig.~1. The typical in-plane center-to-center spacing of neighboring nanoparticles is $d \approx 8nm$ for all samples. From the corresponding diffraction patterns in the insets, it is evident that successive layering retains the very high degree of uniformity and order similar to that achievable with monolayers~\cite{Parthasarathy}. There is no epitaxial registry between layers, rather they appear to be slightly rotated with respect to each other.

\begin{figure}
\begin{center}
\includegraphics [width=3in] {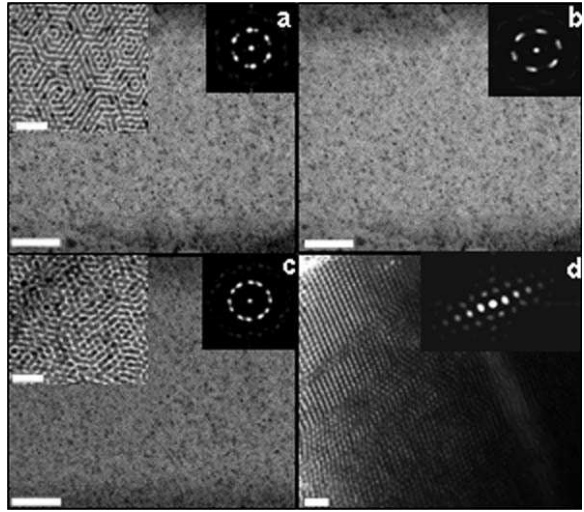}
\caption{Transmission electron micrographs showing the region between the in-plane electrodes for a) bilayers, b) trilayers, c)tetralayers and d) thick films. The darker regions on top and bottom of a-c are the electrodes. The insets on the right sides are diffraction patterns computed by fast Fourier transform. The insets on the left sides of panels a$\&$c are the zoomed-in images. The scale bars correspond to 200nm (a-c) and 40nm (d, all insets). }\label{fig.1}
\end{center}
\end{figure}

\begin{figure*}
\begin{center}
\includegraphics[width=7.0in]{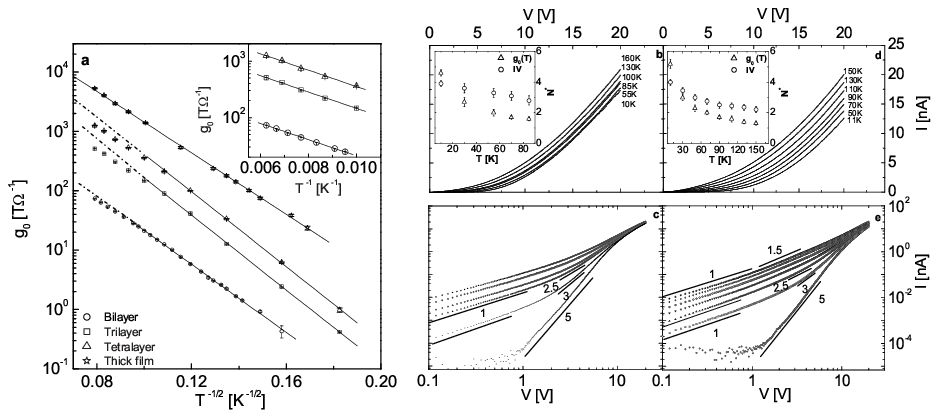}
\caption{a) Zero-bias conductance $g_0$ versus inverse temperature $T^{-1/2}$ for representative multilayer and thick film data.  The thick-film curve has been shifted upward for clarity by multiplying the data by 40. Inset: For the high-temperature range, where the multilayer data in the main panel deviate form the dotted lines, $g_0$ has been replotted as a function of $T^{-1}$, indicating Arrhenius behavior from 100-160K(b-e) Evolution of the $I-V$ characteristics with temperature for bilayers (b,c) and thick films (d,e). Panels (c) and
(e) are log-log plots of the data shown in the plots above them. The straight solid lines are guide to the eye, indicating power law behavior. The noise limit in these measurements was about 25fA. Insets to b$\&$d: Temperature dependence of the hopping distance $N^*$ obtained from $g_0(T)$ and the I-V power-law exponents obtained from panels c$\&$e in the range $2V<V<7V$. }\label{fig.2}
\end{center}
\end{figure*}

Measurements of the zero-bias conductance, $g_0(T)$, as well as the full $I-V$ characteristics exhibit only small differences between bilayers, trilayers, tetralayers and thick films (Fig.2). The bilayer conductance is several order of magnitude larger than in ordered monolayers prepared from the same particles and measured under identical conditions~\cite{Parthasarathy}, but does not increase proportionally when more layers are added. This indicates (i) a strong inter-layer coupling, and (ii) that current flows predominantly through the layers closest to the substrate, consistent with our in-plane electrode configuration. A finite $g_0(T)$ is measurable down to temperatures $T^*$ as low as 25-30K, given the noise limit of about 25fA in our set-up. By contrast, in monolayers  the accessible temperature range for $g_0(T)$ is too small to unambiguously identify its functional dependence because of the significantly higher $T^*$, around 100K and larger.

Figure 2a shows that the zero-bias conductance follows $g_0(T)\sim \exp [-(T_0/T)^{1/2}]$ over the range 30-90K for the multilayers and 30-150K for the thick films.  This is the functional form resulting both from the ES VRH model and the newer approaches based on cotunneling. Fits indicated by the lines give $T_0=(4.00\pm 0.02)\times 10^3 K$ and $(3.00\pm 0.01)\times 10^3K $ for multilayers and thick films, respectively. At temperatures above $ \approx$100K, $g_0(T)$ for the multilayers starts to deviate from the low-temperature behavior (dashed lines) and crosses over to Arrhenius behavior $g_0(T)  \sim \exp [-U/k_BT]$ (Fig.2a, inset) with activation energies $U/k_B \approx 320\pm8K$.

Associating this high-temperature behavior with nearest-neighbor tunneling between particles with randomly offset Fermi levels in an effectively 2D system (because only a few layers contribute to the transport), we can use U$\approx$0.2$E_C$, in analogy to monolayers~\cite{Parthasarathy}. This gives $E_C\approx$1600K. Here $E_C =e^2/4\pi\epsilon\epsilon_0 a$ is the Coulomb charging energy of an individual grain, expressed in terms of grain radius $a$, electron charge, $e$, permittivity of space, $\epsilon_0$ , and dielectric constant of the surrounding medium, $\epsilon$.  $E_C\approx$1600K implies $\epsilon \approx 4$, larger than for monolayers on insulating substrates and in line with the extra screening provided by multilayers. $T_0$ is related to a localization length, $\xi$, through $T_0 = \frac{2.8e^2}{4 \pi \epsilon \epsilon_0 k_B \xi}$~\cite{Efros}. Using the above values for $\epsilon$ and $T_0$, we find $3nm < \xi <4nm$ for both multilayers and thick films, i.e., localization to within one grain. This is an important check as larger values would have implied strongly coupled clusters of grains.

The typical hopping distance, $r^*$, is obtained by maximizing the ES hopping probability $P(r)=\exp[-\frac {r}{\xi}-\frac{e^2} {4\pi \epsilon \epsilon_0 T k_B r}]$ with respect to $r$. This gives $r^* = (\frac{e^2\xi}{4 \pi \epsilon\epsilon_0 k_B T})^{1/2}$. The number of grains, $N^*$, involved in a typical hop is $N^*=r^*/d$, with a center-to-center distance $d \approx 8nm$ between neighboring grains in our samples. At $T=10K$ this leads to $N^*=4$ for multilayers and 4-5 for the thick films. As temperature increases, $N^*$ decreases until $N^*\sim 1$, at which point a crossover to simple activated transport takes place. For the multilayers, we expect this to happen at $T\approx 90-95K$  but for the thick films only above $T \approx 130K$. Both predictions are in excellent agreement with the data in Fig.2a.

This behavior differs from that for doped semiconductors, where a crossover from electron-interaction-dominated (ES-type) to single-electron(Mott-type) VRH~\cite{Mott} is expected with increasing temperature~\cite{Efros} \textit{before} the crossover to Arrhenius behavior takes place. The switching from ES- to Mott-type VRH is associated with a temperature $ T_{cross} = T_M \left(\frac{T_{ES}}{T_M}\right)^{\frac{D+1}{D-1}}$. Here $D$ is the dimensionality of the sample, $T_M= 1/\nu_0\xi^D$ where $\nu_0 = 1/E_C a^D$ is the bare density of states, and $T_{ES}= e^2/(4\pi\epsilon\epsilon_0\xi)$. It follows that $T_{cross} \approx 1400K$ for multilayers (2D) as well as thick films (3D). Such large $T_{cross}$ implies $N^*<<1$, consequently, Mott-type VRH should not be observed in granular metals and quantum dot arrays \cite{Beloborodov,Yakimov}.

A further prediction of the cotunneling scenario connects the exponent of the power law $I-V$ curves to the number of junctions involved. For elastic cotunneling, $I \sim V$, in contrast to the data in Fig.2. However, for inelastic cotunneling through j=$N^*$-1 junctions in a row it can be shown from refs.~\cite{Averin,Geerligs} that
\begin{equation}
\label{I} I_{in} \sim V \left[\frac{g_T}{h/e^2}\right ]^j
\left[\frac{(eV)^2+(k_BT)^2}{E_C^2}\right]^{j-1},
\end{equation}
where $g_T$ is the tunnel conductance of a single junction. The factor $(\frac{g_T}{h/e^2})^j$ explicitly shows that the observation of cotunneling quickly becomes difficult for small $g_T$ and large $j$. Thus, multiple cotunneling is required. For temperatures $E_C>\frac{eV}{N} \gg k_BT$, Eq.1 predicts $I \sim V^{1+2(j-1)}$. From the estimates of $N$ at $T=10K$, above, we expect typical hops of about 4-4.5 grains, i.e., roughly 15 cotunneling events to make it across the electrode gap of about $60d$ in our samples. N=4 or j=3 implies $I \sim V^5$. 

This power law is precisely what emerges in the  lowest temperature $I-V$ characteristics, as soon as our $\approx 25fA$ noise level is exceeded. As Figs. 2c and e show, multilayers and thick films behave similarly  (the $I-V$ characteristics of trilayers and tetralayers are identical in shape to those of bilayers and are not shown here). At elevated $T$ the reduced typical hop distance furthermore leads to smaller $j$ and thus smaller power law exponent in the $I-V$s. As the insets to Figs. 2b$\&$d demonstrate, the number of grains $N^*$ obtained from the typical hopping distance $r^*$ and from the power law exponents of the $I-V$ curves track each other very well. Once $E_C>k_BT \gg\frac{eV}{N}$ the temperature term in Eq.~(\ref{I}) dominates, and we expect $I \sim V$. Essentially linear low-bias portions of the $I-V$s are indeed observed at higher temperatures, and they extend to larger applied bias as $T$ increases. Finally, at temperatures above roughly 100K, where  the analysis of $g_0(T)$ predicts a crossover from ES-like to simple activated behavior, $N=1$ and the exponent in the $I-V$ characteristics approaches unity over the whole low-bias, Coulomb blockade regime. 

For larger applied bias, the Coulomb blockade regime is exceeded and the $I-V$ characteristics will cross over from cotunneling to field-driven behavior.  This is expected for $V >> V_t$, where $V_t$ is the threshold for the Coulomb-blockade regime. At low temperatures, $V_t \approx 0.23NE_C$ for an array of  $N \times N$ grains ~\cite{Parthasarathy}, leading to values around 2V for the data in Fig.2.  Theory ~\cite{Middleton}, simulations ~\cite{Middleton,Reichhardt} and earlier results on highly-ordered monolayers~\cite{Parthasarathy}, predict $I \sim (V-V_t)^\xi$ in this high-bias regime, with exponent $\xi$ close to 2. Our data in Figs. 2c$\&$e is consistent with such cross-over and $\xi \approx 2-2.5$. However, since the cotunneling regime extends significantly above $V_t$ the remaining, high-bias voltage range is too small to reliably extract a value for $\xi$.

These results demonstrate that a picture based on multiple inelastic cotunneling can account for both the temperature dependence of both the zero-bias conductance and the nonlinear $I-V$ characteristics in the  Coulomb blockade regime of large nanoparticle arrays.  It also ties in well with previous results focusing on the behavior at higher bias. The use of  highly ordered arrays assembled from nanoparticles with very tight dispersion , in conjunction with direct TEM imaging, makes it possible to rule out other scenarios \cite{Sheng} that are based on disorder as the mechanism behind the ES-like VRH conductivity. Since details of the individual particle properties, such as their density of states, do not enter in the scenario developed here, we expect qualitatively similar behavior for a wide range of similar systems.

We thank T. Bigioni, K. Elteto, I.Gruzberg and A. Lopatin, for insightful discussions, and M. Constantinides, R. Diamond, Q. Guo. X.-M. Lin. acknowledges support from DOE W-31-109-ENG-38. This work was supported by the UC-ANL Consortium
for Nanoscience Research and by the NSF MRSEC program under DMR-0213745.

\end {document}